\documentclass[prb,twocolumn,superscriptaddress,floats,citeautoscript]{revtex4}

\usepackage{graphicx}
\usepackage{amsmath}
\usepackage{amssymb}
\usepackage{times}
\usepackage{colordvi}
\usepackage{bm}

\newcommand* {\vek}[1]{{\ensuremath{\bm{\mathrm{#1}}}}}

\newcommand{\ket}[1]{\left |  #1 \right \rangle}

\graphicspath{{.}{./EPS/}}

\begin{document}

\title{Land\'e-like formula for the $\bm g$ factors of hole-nanowire subband edges}

\author{D. Csontos}
\affiliation{Institute of Fundamental Sciences and MacDiarmid Institute for Advanced 
Materials and Nanotechnology, Massey University, Private Bag 11~222, Palmerston
North 4442, New Zealand}

\author{U. Z\"ulicke}
\affiliation{Institute of Fundamental Sciences and MacDiarmid Institute for Advanced 
Materials and Nanotechnology, Massey University, Private Bag 11~222, Palmerston
North 4442, New Zealand}

\author{P. Brusheim}
\affiliation{Division of Solid State Physics, Lund University, Box 118, S-22100 Lund,
Sweden}

\author{H.~Q. Xu}
\affiliation{Division of Solid State Physics, Lund University, Box 118, S-22100 Lund,
Sweden}

\date{\today}

\begin{abstract}

We have analyzed theoretically the Zeeman splitting of hole-quantum-wire subband
edges. As is typical for any bound state, their $g$ factor depends on both an intrinsic
$g$ factor of the material and an additional contribution arising from a finite
bound-state orbital angular momentum. We discuss the
quantum-confinement-induced interplay between bulk-material and orbital effects,
which is nontrivial due to the presence of strong spin-orbit coupling. A compact
analytical formula is provided that elucidates this interplay and can be useful for
predicting Zeeman splitting in generic hole-wire geometries.

\end{abstract}


\maketitle

The interaction of a particle's magnetic moment with an external magnetic field
$\vek B$ typically results in the lifting of degeneracies in its energy spectrum. The
study of such Zeeman splittings provides important clues about microscopic
properties of a quantum system~\cite{vanVleck}. In vacuum or the bulk of a solid,
the Zeeman splitting can be understood as arising from the coupling of a (quasi-)free
particle's intrinsic angular momentum (i.e., spin) to $\vek B$. However, in general, 
orbital motion in a quantum bound state is associated with a magnetic moment
that interacts with the magnetic field and, thus, also contributes to Zeeman splitting
and the $g$ factor. In the presence of spin-orbit coupling, the interplay between
orbital and intrinsic (spin) contributions to $g$ can be nontrivial~\cite{vanVleck}.

Recently, \textit{p}-type semiconductor nanostructures have become available as
interesting laboratories to explore the interplay between quantum confinement and
the magnetic moment of spin-3/2 holes~\cite{roland:prl:00,haug:prl:06,
uz:prl:06,li:nlett:07,pryor:prl:06,pryor:prb:07,sheng:prb:07,kai:apl:07}. For example, 
strong spin-orbit coupling present in the upper-most valence band of typical
semiconductors results in an energy splitting~\cite{sherman:pla:88} between
heavy-hole (HH) and light-hole (LH) quantum-well subband
edges~\cite{HHandLHnote}. This HH-LH splitting turns out to freeze the HH
spin-projection axis to be perpendicular to the quantum well, thus preventing its
reorientation by an in-plane magnetic field and, therefore, suppressing Zeeman
splitting~\cite{roland:prl:00,rolandbook}. In addition to HH-LH {\em splitting\/},
confinement typically induces a {\em mixing\/} of HH and LH states in quantum
wires~\cite{bastardrev, sercel:prb:90, uz:prb:07b, zhu:nlett:07}, even at subband 
edges. Both HH-LH mixing and splitting strongly affect the physical properties of
hole states.

We have previously shown~\cite{uz:prb:07b,uz:apl:08} how HH-LH mixing modifies
the Zeeman splitting of hole-wire subband-edges when the bulk-material hole $g$
factor $\kappa$ is large and can be assumed to dominate over all orbital
magnetic-field effects~\cite{PrevWorkNote}. Here we complement this analysis of
Zeeman splitting in hole quantum wires, focusing on the confinement-induced
interplay between bulk-material and orbital bound-state contributions to the $g$
factor. Bulk-material and orbital contributions turn out to vary strongly between
different quantum-wire subband edges, in both their magnitude and sign. Depending
on their relative sign, the two contributions can either enhance or suppress each
other, thus yielding large variations in sign and magnitude of the total $g$ factor. We
provide a formula from which the origins of the observed spin-splitting characteristics
can be explained and which may also be useful to predict Zeeman splittings in
generic hole-wire geometries. For cylindrical hole wires, we also find a Land\'e-like
formula, which explicitly expresses the subband-edge $g$ factor in terms of a new
total angular momentum and spin-3/2 tensor invariants~\cite{roland:prb:04}, the latter
resulting from a multipole expansion of the spin-3/2 density matrix. It is an interesting
feature of spin-3/2 systems that, in addition to a monopole and dipole moment, which
are related to the density and polarization, respectively, a quadrupole and an
octupole moment can exist, which have no equivalents in spin-1/2 systems. Orbital
contributions turn out to give rise to a dependence of the hole-wire $g$ factors on
spin-3/2 quadrupole and octupole moments.

\begin{figure}[b]
\includegraphics[width=2.6in]{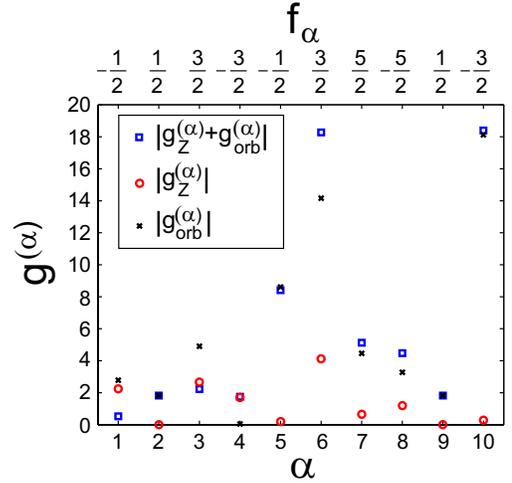}
\caption{\label{fig:1}
(Color online) Absolute values of orbital and bulk-material Zeeman contributions,
$g_{\text{orb}}^{(\alpha)}$ and $g_{\text{Z}}^{(\alpha)}$, to the total $g$-factor,
$g_{\text{tot}}^{(\alpha)}=g_{\text{Z}}^{(\alpha)}+g_{\text{orb}}^{(\alpha)}$, for a
cylindrical hole nanowire defined by a hard-wall confinement in GaAs (i.e., a material 
with Luttinger parameters $\gamma_1=6.98$, $\gamma_{\text{s}}=2.58$, and
$\kappa=1.2$; see text). Values are shown for the highest-in-energy subband edges
enumerated by $\alpha$. $f_{\alpha}$ is the quantum number of the total angular
momentum operator $\hat{F}_{z}=\hat{J}_{z}+\hat{L}_{z}$. The magnetic field is
assumed to be parallel to the wire ($z$) axis.}
\end{figure}
\begin{table*}[t]
\caption{$g$-factor contributions for cylindrical-hole-nanowire subband edges in
GaAs, with cubic corrections neglected.}
\begin{tabular}{|c|rrrrrrrrrr|}
\hline
$\alpha,f_{\alpha}$ & 1, $-\frac{1}{2}$ & 2, $\frac{1}{2}$ & 3, $\frac{3}{2}$  &
4, $-\frac{3}{2}$  & 5, $-\frac{1}{2}$ & 6, $\frac{3}{2}$ & 7, $\frac{5}{2}$  &
8, $-\frac{5}{2}$ & 9, $\frac{1}{2}$ & 10, $-\frac{3}{2}$ \\[0.1cm] \hline
$g_{\text{tot}}^{(\alpha)}$ &  -0.52 & -1.82 & 2.23 & 1.75  & 8.44 & -18.23  & -5.12 &
4.48   & -1.82 & 18.34 \\
$g_{\text{Z}}^{(\alpha)}$ &  2.26 & 0.00 & -2.68  & 1.70  &  -0.19 & -4.12   & -0.66 & 1.20  & 0.00 & 0.28 \\
$g_{\text{orb},\text{diag}}^{(\alpha)}$&0.55 & -1.82 & -8.27  & 12.30 & 10.32 & -5.66 
& -24.1 & 24.9  &  -1.82 & 19.5 \\
$g_{\text{orb},\text{mix}}^{(\alpha)}$& -3.33 & 0.00 &13.18 & -12.25  & -1.69  & -8.45 
& 19.64 & -21.62  & 0.00 & -1.44 \\[0.05cm] \hline
\end{tabular}
\label{table:1}
\end{table*}
Figure \ref{fig:1} shows our calculated absolute values of bulk-material and orbital
contributions $\vert g_{\text{Z}}^{(\alpha)}\vert$, $\vert g_{\text{orb}}^{(\alpha)}\vert$
to the total $g$-factor $\vert g_{\text{tot}}^{(\alpha)}\vert=\vert g_{\text{Z}}^{(\alpha)}
+g_{\text{orb}}^{(\alpha)}\vert$, respectively, for cylindrical-hole-wire subband edges.
The integer $\alpha$ enumerates wire subbands, starting with the highest. The
magnetic field is assumed to be applied parallel to the wire, which is aligned with the
$z$-coordinate axis, and we used materials parameters characteristic for
GaAs~\cite{vurg:jap:01} but with band warping neglected. Several observations can
be made. i)~The orbital contributions generally have a magnitude of the same order
as that of the corresponding bulk-material ones. In some instances, such as for levels
$\alpha=5,6$ and 10, they greatly exceed the bulk-material contributions, giving rise
to very large spin splittings. ii)~The bulk-material and orbital contributions fluctuate in
sign and magnitude between different subbands. iii)~The two contributions can
variably enhance or suppress each other, thus rendering large variations in sign and
magnitude of the total $g$-factor $g_{\text{tot}}^{(\alpha)}$ as a function of subband
index $\alpha$. Further below we discuss how the observed behavior results from
HH-LH mixing.

Before describing details of our calculational method and giving further results, we
discuss the general form of the various $g$-factor contributions. As the hole-wire
subband edges are well-separated in energy, it is possible to calculate their Zeeman
splitting analytically within a perturbative approach that becomes exact in the limit of 
$\vek{B} \equiv B_z\,\vek{\hat z}\rightarrow \vek{0}$. We find the following general
expression $g_{\text{tot}}^{(\alpha)}= g_{\text{Z}}^{(\alpha)} +
g_{\text{orb},\text{diag}}^{(\alpha)} + g_{\text{orb},\text{mix}}^{(\alpha)}+g_{\text{orb},
\text{cub}}^{(\alpha)}$. The first three terms are independent of the wire's direction
with respect to crystallographic axes, neglect band warping, and have the following
form:
\begin{subequations}\label{gFactTerms}
\begin{eqnarray} \label{gFactZee}
g_{\text{Z}}^{(\alpha)} &=& -4\kappa \langle \hat J_z \rangle_\alpha \quad , \\
\label{gFactDiag}
g_{\text{orb},\text{diag}}^{(\alpha)} &=& - 2 \gamma_{1} \langle \hat{L}_{z}\,
\openone_{4\times 4} \rangle_\alpha  
- 2 \gamma_{s} \left\langle\left ( \hat{J}_{z}^{2}-\frac{5}{4} \openone_{4\times 4}
\right ) \hat{L}_{z} \right\rangle_\alpha , \\ \label{gFactMix}
g_{\text{orb},\text{mix}}^{(\alpha)} &=&  - 2 \gamma_{s} \left\langle i \hat{x}_{-}
\hat{k}_{-} \hat{J}_{+}^{2} - i \hat{x}_{+}\hat{k}_{+}\hat{J}_{-}^{2} \right\rangle_\alpha
\, .
\end{eqnarray}
\end{subequations}
Here $\langle\dots\rangle_\alpha$ denotes the expectation value of the operator
inside the angular brackets taken for the positive-helicity
eigenstate~\cite{sercel:prb:90,helicity} of wire subband edge $\alpha$ obtained in
zero magnetic field. $\kappa$ is the bulk-hole $g$-factor,~\cite{luttham2} $\hat{L}_{z}
=x\hat{k}_{y}-y\hat{k}_{x}$, $\hat{x}_{\pm}=x\pm iy$, $\hat{k}_{\pm}=\hat{k}_{x}\pm i
\hat{k}_{y}$, $\hat{J}_{\pm}=(\hat{J}_{x}\pm i \hat{J}_{y})/\sqrt{2}$, $\hat{k}_{i}$ are
orbital linear momenta, $\hat{J}_{i}$ are spin-3/2 angular-momentum operators, and
$\gamma_{s}=(2\gamma_{2}+3\gamma_{3})/5$ in terms of Luttinger
parameters~\cite{luttham2,lip:prl:70}. The band-warping correction
$g_{\text{orb},\text{cub}}^{(\alpha)}$ depends on the wire direction~\cite{cubicterm}.
For now, we continue neglecting this term in our calculations, postponing a discussion
of its effect on our results to the end of the paper.

As previously discussed, it is a general characteristic of hole nanowire subband 
edges that they are mixtures of HH and LH
characters~\cite{bastardrev, sercel:prb:90, uz:prb:07b}. Thus, all of the individual 
terms in the expression of $g_{\text{tot}}^{(\alpha)}$  are affected by HH-LH mixing, 
because the expectation value is taken with respect to a HH-LH mixed nanowire 
subband-edge state. It was previously shown that the bulk-material contribution
$g_{\text{Z}}^{(\alpha)}$ fluctuates strongly between different subband edges as a 
result thereof~\cite{uz:prb:07b,uz:apl:08}.

The orbital contributions $g_{\text{orb},\text{diag}}^{(\alpha)}$ and $g_{\text{orb},
\text{mix}}^{(\alpha)}$ show similar behaviour, as can be seen from their values 
given in Table~\ref{table:1}. Interestingly, with only few exceptions (e.g., the level 
with $\alpha=6$), these two terms appear to counteract each other, sometimes quite 
dramatically. (See e.g., levels with $\alpha =4$ and 8.) $g_{\text{orb},
\text{diag}}^{(\alpha)}$ has a straightforward interpretation in that it represents the
na{\"\i}ve coupling of a bound-state orbital magnetic moment to the external 
magnetic field, with different gyromagnetic ratio for HH and LH states because of 
their different effective masses. The term $g_{\text{orb},\text{mix}}^{(\alpha)}$ arises
because the spin-orbit coupling in the valence band induces linear-in-$B$ terms that
couple HH and LH amplitudes directly. Thus it represents a HH-LH mixing effect on
top of the mixing already present in the spin-3/2 subband-edge states.

The total $g$ factor $g_{\text{tot}}^{(\alpha)}$ is a sum of bulk-material and orbital
terms. As seen in Table~\ref{table:1}, these different terms can vary substantially in
magnitude and sign between different subband edges $\alpha$ as a result of the
HH-LH mixing. More importantly, the relative sign between the various contributions
is different between the subband edges. Thus, the bulk-material and orbital terms
can varyingly enhance or suppress each other's contribution to the total $g$ factor
depending on the subband index $\alpha$.

We now proceed to discussing in detail the theoretical formalism for calculating the
total $g$-factor and its individual contributions given in
Eqs.~(\ref{gFactZee})--(\ref{gFactMix}). Subsequently, we will show that their basic
characteristics discussed above are universal for quasi-1D hole systems by
considering the influence of reduced symmetry of the confining potential. We will
also analyze the effect of (so far neglected) cubic corrections. 

Our starting point is the Luttinger Hamiltonian\cite{luttham2} in the spherical 
approximation\cite{lip:prl:70}
\begin{equation}
\label{HLuttinger}
H_{\text{L}} =  -\frac{\gamma_{1}}{2}\,\hat{k}^{2}\,{\openone}_{4\times 4}+\gamma_{s}
\left [\left (\hat{\vek{k}}\cdot \hat{\vek{J}} \right )^{2} - \frac{5}{4}\,\hat{k}^{2}\,
{\openone}_{4\times 4} \right ]  ~~.
\end{equation}
In Eq.\ (\ref{HLuttinger}), we used atomic units. The spin-orbit-coupling term
proportional to $\gamma_{\text{s}}$ is the origin of HH-LH splitting and mixing,
giving rise to the peculiar spin properties of confined holes. In the following, we 
assume $\hat{k}_{z}=0$ because we will focus solely on the properties of quasi-1D
subband edges~\cite{IAcaveat}. The bulk-hole Zeeman interaction with a magnetic
field applied along the wire ($z$) direction is given by $H_{\text{Z}}=-2\kappa
\mu_{\text{B}}B_{z} \hat{J}_{z}$. (We neglect the small anisotropic
part~\cite{rolandbook} of the Zeeman splitting in the bulk valence band.) This term
is the origin of the contribution $g_{\text{Z}}^{(\alpha)}$ to the hole-wire $g$ factor.
The orbital effect of the magnetic field is included by replacing $\hat{\vek{k}}
\rightarrow \hat{\vek{k}} +{\vek{A}}$, with the symmetric gauge ${\vek{A}}=
(-y/2,x/2,0)B_{z}$. As we will only consider the leading (linear-in-$B_z$) contribution
to the Zeeman splitting, we neglect orbital terms of higher than linear order in
$B_{z}$. The additional Hamiltonian, $H_{\text{orb}}$, which incorporates the effects
due to the interaction between the holes' orbital degrees of freedom and the magnetic
field, is then given by
\begin{widetext}
\begin{equation}
H_{\text{orb}} = - \mu_{\text{B}} B_{z} \left\{ \left [\gamma_{1} \openone_{4\times 4} 
+ \gamma_{s} \left ( \hat{J}_{z}^{2}-\frac{5}{4}\openone_{4\times 4}\right ) \right ]
\hat{L}_{z} + i \gamma_{s} \left ( \hat{x}_{-}\hat{k}_{-}\hat{J}_{+}^{2}-\hat{x}_{+}
\hat{k}_{+}\hat{J}_{-}^{2} \right )  \right\} ~~.
\label{Horb}
\end{equation}
\end{widetext}
The first term, proportional to $\hat{L}_{z}$ and diagonal in spin-3/2 space, gives
rise to the %diagonal orbital
contribution $g_{\text{orb},\text{diag}}^{(\alpha)}$. The
second term couples HH and LH states and leads to
$g_{\text{orb},\text{mix}}^{(\alpha)}$.

For $\hat k_z =0$, the Hamiltonian $H_B = H_{\text{L}}+H_{\text{Z}}+H_{\text{orb}}$ 
turns out to be block-diagonal in spin-3/2 space such that subspaces for helicity
$\pm$ do not couple\cite{sercel:prb:90,helicity}. Subband energies $E_{\alpha 
\pm}$ within the $\pm$ subspaces are found by diagonalising corresponding blocks
in $H_B+V(x,y)$, where $V(x,y)$ is the confining potential defining the wire. For any
given $\alpha$, eigenstates $\ket{\alpha\pm}$ with energies $E_{\alpha\pm}$ are
found to be degenerate at zero magnetic field, but these split for $B_z>0$. We define
the  $g$ factor for a hole-wire subband edge with index $\alpha$ according
to~\cite{gDefComment}  $g_{\text{tot}}^{(\alpha)}=\lim_{B_z\to 0} [E_{\alpha+}(B_z) - 
E_{\alpha -}(B_z)])/(\mu_{B} B_{z})$. Because of the finite energy separation of
subband edges for different $\alpha$, this definition turns out to be equivalent to the
perturbative (to first order in $B_z$) result
\begin{eqnarray}
g_{\text{tot}}^{(\alpha)} &=& \frac{\langle H_{\text{Z}}+H_{\text{orb}} \rangle_{\alpha 
+} - \langle H_{\text{Z}}+H_{\text{orb}} \rangle_{\alpha -}}{\mu_{\text{B}}
  B_{z}} \,\, \nonumber  \\
&\equiv& \frac{2 \langle H_{\text{Z}}+H_{\text{orb}} \rangle_{\alpha +}}{\mu_{\text{B}}
B_{z}} ~~,
\label{gpert}
\end{eqnarray}
where the expectation values are taken with respect to the wire eigenstates that 
diagonalize the zero-field, quantum-confined Luttinger Hamiltonian $H_{\text{L}}
+V(x,y)$. 

An interesting case arises for hole wires with cylindrical cross-section. There, 
subband-edge bound states are also eigenstates of a new total angular momentum
$\hat F_z \equiv \hat J_z + \hat L_z$~\cite{sercel:prb:90,uz:prb:07b,uz:apl:08}. This
makes it possible to rewrite Eqs.~(\ref{gFactTerms}) in terms of the quantum
number $f_\alpha$ associated with $\hat F_z$, as well as expectation values of
spin-3/2 operators, thereby essentially eliminating any dependence on the orbital
angular momentum $\hat L_z$.

In the spherical approximation, we find $g_{\text{tot},\text{cyl}}^{(\alpha)}=
g_{\text{d},\text{cyl}}^{(\alpha)}+g_{\text{q},\text{cyl}}^{(\alpha)}+g_{\text{o},
\text{cyl}}^{(\alpha)}+g_{\text{r},\text{cyl}}^{(\alpha)}$, where
\begin{subequations}\label{cylZeemTerms}
\begin{eqnarray}
g_{\text{d},\text{cyl}}^{(\alpha)} &=& -2 \gamma_1 f_\alpha - ( 4\kappa - 2\gamma_1 ) \left \langle \hat J_z \right\rangle_\alpha \, , \\
g_{\text{q},\text{cyl}}^{(\alpha)} &=& -2 \gamma_{\text{s}} f_\alpha \left \langle \hat
J_z^2 - \frac{5}{4} \openone_{4\times 4} + \hat J_+^2 + \hat J_-^2  \right
\rangle_\alpha \, , \\
g_{\text{o},\text{cyl}}^{(\alpha)} &=& 2 \gamma_{\text{s}} \left\langle \left[ \hat J_z^2 -
\frac{5}{4} \openone_{4\times 4} + \hat J_+^2 + \hat J_-^2 \right] \hat J_z \right
\rangle_\alpha \, , \\
g_{\text{r},\text{cyl}}^{(\alpha)} &=& -2 \gamma_{\text{s}} \left\langle \left[ \hat J_+^2 - 
\hat J_-^2 \right] r \partial_r \right\rangle_\alpha \, .
\end{eqnarray}
\end{subequations}
Here $r$ is the radial coordinate in the $xy$ plane, and expectation values are to be
taken in the radial spinor wave functions that are eigenfunctions of the canonically
transformed Hamiltonian $e^{i \hat J_z \varphi} [ H_{\text{L}} + V(r) ]e^{-i \hat J_z
\varphi}$\cite{uz:prb:07b}. 

In Eqs.~(\ref{cylZeemTerms}a-c), expectation values are taken of spherical tensor
operators associated with multipole invariants of the spin-3/2 density
matrix~\cite{roland:prb:04}. The contribution $g_{\text{d},\text{cyl}}$ depends on the 
dipole moment, or polarization, of the hole-wire bound state. It contains the
previously analyzed~\cite{uz:prb:07b,uz:apl:08} bulk-material contribution that is
proportional to $\kappa$. Interestingly, expectation values of tensor components for
spin-3/2 quadrupole and octupole moments appear in the additional terms
$g_{\text{q},\text{cyl}}$ and $g_{\text{o},\text{cyl}}$, respectively. 
Such quadrupole and octupole moments are a unique feature of spin-3/2 systems 
and have no equivalent in the spin-1/2 realm. Situations can arise in which the 
dipole moment (polarization) is very small, but higher moments can be substantial. 
For example, a magnetic field applied parallel to a two-dimensional hole system can 
induce a large octupole moment, whilst the dipole moment associated with the 
polarization is vanishing\cite{roland:prb:04}. For the hole nanowires, the appearance
of higher moments in the $g$ factor are entirely due to orbital effects, which is
signified by their dependence on the Luttinger parameters. The remainder term
$g_{\text{r},\text{cyl}}$ is of purely orbital origin but does not appear to have a
straightforward relation to pure spin-3/2 tensors.

\begin{figure}[t]
\includegraphics[width=2.6in]{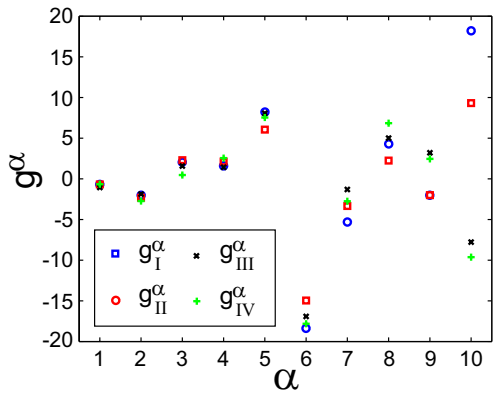} 
\caption{(Color online) Total $g$-factors for hole wire subband edges labelled by
index $\alpha =1\dots 10$, calculated using a hierarchy of approximations described 
in the text. A close agreement between the different results for low-index subbands
is observed, while corresponding numbers show an increasing spread for
higher-index subband edges.
\label{fig:2}}
\end{figure}
To verify the validity of our approximations and the use of the perturbation approach,
we have performed calculations taking into account the effects of lower symmetries 
of the confining potential, as well as the underlying crystal, and compared these 
results with numerical diagonalization of the full Luttinger Hamiltonian including 
band warping~\cite{luttham2,lip:prl:70,cubicterm} Our results are shown in
Fig.~\ref{fig:2}, where we plot the total $g$-factors for the ten highest wire subband
edges with index $\alpha=1\dots 10$, calculated using a hierarchy of four different
approximations: {\bf I} Perturbation theory for $H_{\text{orb}}+H_{\text{Z}}$ using 
unperturbed states that diagonalize the Hamiltonian $H_{L}+V(x,y)$ for a cylindrical 
hard-wall potential. (These are the same results as shown in Fig.~\ref{fig:1} and 
Table \ref{table:1}). {\bf II} Same as {\bf I}, except that the wire is defined by a
hard-wall potential with {\em square\/} crossection, thus reducing the symmetry of
the confining potential. {\bf III} Same as {\bf II}, but the unperturbed states are
obtained by diagonalizing the zreo-field Luttinger Hamiltonian including band-warping
terms, thus taking into account the lower (cubic) crystal symmetry~\cite{luttham2}.
(For this case, the wire is assumed to be parallel to the [001] direction.) {\bf IV}
Results obtained from full numerical diagonalization of the Luttinger Hamiltonian with
finite magnetic field, including band warping, for a square cross-section wire parallel
to the [001] direction~\cite{cubicterm}. Comparison between these different cases
shows good agreement for states with low subband index $\alpha$. For higher
$\alpha$, however, the effects of reduced symmetry are more pronounced, and there
appears an increased spread between the $g$-factors calculated for different
approximations. We expect that these will become even more important for wires
aligned with crystallographic directions other than the high-symmetry [001] direction
considered in our calculations.

In conclusion, we have presented a comprehensive and general~\cite{interactComm}
formalism for the description of the Zeeman splitting of hole-quantum-wire subband
edges and applied it to hole wires subject to a magnetic field applied parallel to the 
wire axis. We have elucidated the role of HH-LH mixing in hole nanowire systems, 
as well as the intricate quantum-confinement-induced interplay between the
bulk-material and orbital contributions to the total $g$ factors. The latter may be used
for tailoring  of hole spin splittings by tuning the relative importance of the two
contributions, e.g., by electrostatic confinement engineering in $p$-type quantum
point contacts~\cite{oleh:apl:06} or nanowires~\cite{li:nlett:07}.

DC acknowledges partial support from the Massey University Research Fund and
thanks the Division of Solid State Physics/Nanometer Structure Consortium, Lund
University, Sweden, for their hospitality, and the Swedish Research Council (VR) for
financial support during the part of the work that was completed in Sweden.

%\bibliography{general,myself,mesophys,spintronics,spinorbit}\end{document}

\begin{thebibliography}{20}
\bibitem{HHandLHnote}
HH and LH states are spin-3/2-projection eigenstates with eigenvalues $\pm 3/2$
and $\pm 1/2$, respectively.
\bibitem{cubicterm}
The cubic-symmetry term that accounts for band-warping effects depends on the 
wire's direction with respect to crystallographic axes~\cite{luttham2}. For a wire
oriented parallel to the [001] direction, it reads
\begin{eqnarray*}
&& g_{\text{orb},\text{cub}}^{(\alpha)} = - \frac{\gamma_2-\gamma_3}{20} \left[ 12
\left \langle \left( \hat{J}_{z}^{2}-\frac{5}{4}\openone_{4\times 4} \right ) \hat{L}_{z}\right \rangle_\alpha \right. \\ && \left. - 5 \left\langle i \hat x_+ \hat k_+ \hat J_+^2 -
i \hat x_- \hat k_- \hat J_-^2 \right\rangle_\alpha - \left\langle i \hat x_+ \hat k_+
\hat J_-^2 - i \hat x_- \hat k_- \hat J_+^2 \right\rangle_\alpha\right] .
\end{eqnarray*}
\bibitem{IAcaveat}
In this work, we neglect inversion-asymmetry-induced spin-orbit coupling, which could
shift quasi-1D subband edges to finite $k_z$.
\bibitem{gDefComment}
Taking the limit $B_z\to 0$ circumvents ambiguities in the definition of $g$ factors
associated with the, in general, nonlinear magnetic-field dependence of spin
splitting. For wires with cylindrical or square cross-section, our obtained values for
subband-edge $g$ factors are then universal, i.e., independent of wire width.
\end{thebibliography}

\end{document}